\documentstyle [12pt,epsf] {article}
\begin{document}
\setcounter{page}{1}
\def\theequation{\arabic{section}.\arabic{equation}}
\def\theequation{\thesection.\arabic{equation}}
\setcounter{section}{0}

\title{ Quantum and string shape fluctuations \\ in the dual Monopole
Nambu--Jona--Lasinio model with dual Dirac strings\thanks{Supported by the
Fonds zur F\"orderung der wissenschaftlichen Forschung, Project P12495-TPH.}}

\author{M. Faber\thanks{E--mail: faber@kph.tuwien.ac.at, Tel.:
+43--1--58801--5598, Fax: +43--1--5864203} ,  A. N. Ivanov\thanks{E--mail:
ivanov@kph.tuwien.ac.at, Tel.: +43--1--58801--5598, Fax:
+43--1--5864203}~{\Large{$^{\P}$}} , A. M\"uller\thanks{E--mail:
mueller@kph.tuwien.ac.at, Tel.: +43--1--58801--5590, Fax: +43--1--5864203} ,
\\ N. I. Troitskaya\thanks{Permanent Address:
State Technical University, Department of Theoretical
Physics, 195251 St. Petersburg, Russian Federation} ~and~  M.
Zach\thanks{E--mail: zach@kph.tuwien.ac.at, Tel.: +43--1--58801--5502,
 Fax: +43--1--5864203}}

\date{}

\maketitle

\begin{center}
{\it Institut f\"ur Kernphysik, Technische Universit\"at Wien, \\
Wiedner Hauptstr. 8-10, A-1040 Vienna, Austria}
\end{center}

\vskip1.0truecm
\begin{center}
\begin{abstract}
The magnetic monopole condensate is calculated in the dual Monopole
Nambu--Jona--Lasinio model with dual Dirac strings suggested in Refs.[1,2]
as
 a functional of the dual Dirac string shape. The calculation is carried out
 in the tree approximation in the scalar monopole--antimonopole collective
excitation field. The integration over quantum fluctuations of the
dual--vector monopole--antimonopole collective excitation field around the
Abrikosov flux line and string shape fluctuations are performed explicitly.
We claim that
there are important contributions of quantum and  string shape fluctuations
to the magnetic monopole condensate.
\end{abstract}
\end{center}

\newpage

\section{Introduction}
\setcounter{equation}{0}

In Refs.[1,2] there has been suggested the dual Monopole
Nambu--Jona--Lasinio (MNJL) model with dual Dirac strings as a continuum
analogy of Compact Quantum Electrodynamics (CQED) which is defined for
lattices as nonlinear $U\,(1)$
gauge theory. It has a confining phase like QCD [3] and realizes  confinement
of ``color'' electric charges. Thereby, the investigation of CQED should help
us to understand quark confinement. As has been shown in Ref.[4] the
nonperturbative vacuum of CQED behaves like an effective dual
superconductor with magnetic monopoles. Due to magnetic monopoles the
electric flux between quarks rearranges and looks like the field produced
by a dual Dirac string.
As a result quarks interact via a linearly rising potential [5,6] that
realizes quark confinement [7,8] and spontaneous breaking of chiral symmetry
 (SB$\chi$S) [8].

The NJL model [9] can be regarded as some kind of relativistic extension of
the BCS (Bardeen--Cooper--Schrieffer) theory of superconductivity [10]. It
also possesses a nonperturbative vacuum with a ground state of the same kind
as in a superconductor in the
superconducting phase. The latter has been the promoting idea of Refs.[1,2]
to put the NJL--model to the foundation of a continuum space--time model
realizing non--perturbative phenomena of CQED.

The MNJL--model is based on the Lagrangian, invariant under ``color''
magnetic $U(1)$ symmetry, of massless magnetic monopoles, self--coupled
through local four--monopole interaction [1,2]:
\begin{eqnarray}\label{label1.1}
\hspace{-0.5in}&&{\cal L}(x) = \bar{\chi}(x) i \gamma^{\mu} \partial_{\mu}
\chi(x) + G [\bar{\chi}(x) \chi(x)]^2 - G_1
[\bar{\chi}(x)\gamma_{\mu}\chi(x)][\bar{\chi}(x)\gamma^{\mu}\chi(x)],
\end{eqnarray}
where $\chi(x)$ is a massless magnetic ``color''  monopole field, $G$ and
$G_1$ are positive phenomenological constants responsible for the magnetic
monopole condensation and the dual--``color'' vector field mass,
respectively.

The magnetic monopole condensation accompanies the creation of massive
magnetic monopoles $\chi_{M}(x)$ with  mass $M$,
$\bar{\chi}\chi$--collective excitations with quantum numbers of scalar
Higgs meson field $\sigma$ with  mass
 $M_{\sigma} = 2\,M$ and a massive dual--vector field $C_{\mu}$ with mass
$M_C$ defined as [1,2]:
\begin{eqnarray}\label{label1.2}
M^2_C = \frac{g^2}{2 G_1}-\frac{g^2}{8\pi^2}[J_1(M) + M^2 J_2(M)],
\end{eqnarray}
where $J_1(M)$ and $J_2(M)$ are quadratically and logarithmically divergent
integrals [1,2]
\begin{eqnarray}\label{label1.3}
\hspace{-0.2in}J_1(M)&=&\int\frac{d^4k}{\pi^2i}\frac{1}{M^2 - k^2} =
 \Lambda^2 - M^2{\ell n}\Bigg(1 + \frac{\Lambda^2}{M^2}\Bigg),
\nonumber\\
\hspace{-0.2in}J_2(M)&=&\int \frac{d^4k}{\pi^2i}\frac{1}{(M^2 - k^2)^2}
 = {\ell n}\Bigg(1 + \frac{\Lambda^2}{M^2}\Bigg) -
\frac{\Lambda^2}{M^2 + \Lambda^2}.
\end{eqnarray}

Here $\Lambda$ is the ultra--violet cut--off.
The mass of the massive magnetic monopole field $\chi_M(x)$ obeys the
gap--equation [1,2]:
\begin{eqnarray}\label{label1.4}
M = -2 G <\bar{\chi}(0) \chi(0)> =\frac{GM}{2\pi^2} J_1(M).
\end{eqnarray}
After the integration over magnetic monopole degrees of freedom the
effective Lagrangian containing quarks, antiquarks and the fields of scalar
$\sigma$
 and dual--vector $C_{\mu}$ collective excitations reads
\begin{eqnarray}\label{label1.5}
{\cal L}_{\rm eff}(x) &=&\frac{1}{4} F_{\mu\nu}(x) F^{\mu\nu}(x) +
\frac{1}{2} M^2_C C_{\mu}(x) C^{\mu}(x) \nonumber\\
&+&\frac{1}{2} \partial_{\mu} \sigma(x) \partial^{\mu} \sigma(x) -
\frac{1}{2} M^2_{\sigma} \sigma^2(x)\Bigg[1 + \kappa
\frac{\sigma(x)}{M_{\sigma}}\Bigg]^2 \nonumber\\
&+&{\cal L}_{\rm free~quark}(x),
\end{eqnarray}
The coupling constants $g$ and $\kappa$ are related by the constraint
\begin{eqnarray}\label{label1.6}
\frac{g^2}{12\pi^2}J_2(M)=\frac{\kappa^2}{8\pi^2}\,J_2(M) = 1
\end{eqnarray}
or $\kappa^2=2g^2/3$ [1,2].

${\cal L}_{\rm free~quark}(x)$ is the kinetic term for the quark and
antiquark,

\begin{eqnarray}\label{label1.7}
\hspace{-0.2in}{\cal L}_{\rm free~quark}(x) = - \sum_{i=q,\bar{q}} m_i
 \int d{\tau} \Bigg(\frac{d X^{\mu}_i(\tau)}{d\tau} \frac{d
X^{\nu}_i(\tau)}{d\tau} g_{\mu\nu} \Bigg)^{1/2} \delta^{(4)}
(x - X_i(\tau)).
\end{eqnarray}
We consider quark and antiquark as classical point--like particles with
 masses $m_q = m_{\bar{q}} = m$, electric charges $Q_q = - Q_{\bar{q}} = Q$,
 and trajectories $X^{\nu}_q(\tau)$ and $X^{\nu}_{\bar{q}}(\tau)$,
respectively. The field strength $F^{\mu\nu}(x)$ is defined [1,2] as
$F^{\mu\nu}(x) = {\cal E}^{\mu\nu}(x) - {^*}dC^{\mu\nu}(x)$, where
$dC^{\mu\nu}(x) = \partial^{\mu} C^{\nu} (x) - \partial^{\nu} C^{\mu}(x)$,
and ${^*}dC^{\mu\nu}(x)$ is the dual version, i.e.,
${^*}dC^{\mu\nu}(x) = \frac{1}{2} \varepsilon^{\mu\nu\alpha\beta}
dC_{\alpha\beta}(x)\,(\varepsilon^{0123} = 1)$.

The ``color'' electric field strength  ${\cal E}^{\mu\nu}(x)$, induced by a
dual Dirac string, is defined following [1,2] as
\begin{eqnarray}\label{label1.8}
\hspace{-0.2in}&&{\cal E}^{\mu\nu}(x) = Q \int\!\!\!\int  d\tau d \sigma
\Bigg(\frac{\partial X^{\mu}}{\partial \tau}
\frac{\partial X^{\nu}}{\partial \sigma} -
\frac{\partial X^{\nu}}{\partial \tau}
 \frac{\partial X^{\mu}}{\partial \sigma}\Bigg)\delta^{(4)}(x - X),
\end{eqnarray}
where $X^{\mu} = X^{\mu} (\tau,\sigma)$ represents the position of a point
on the world sheet swept by the string. The sheet is parametrized by the
internal coordinates  ${-\infty} < \tau < {\infty}$ and $0 \le \sigma \le
\pi$, so that $X^{\mu} (\tau, 0) = X^{\mu}_{ - Q}(\tau)$ and
$X^{\mu}(\tau,\pi) = X^{\mu}_Q(\tau)$ represent the world lines of an
antiquark and a quark [1,2,5]. Within the definition Eq.(\ref{label1.8})
the tensor field ${\cal E}^{\mu\nu}(x)$ satisfies identically the equation
of motion,
$\partial_{\mu} F^{\mu\nu}(x) = J^{\nu}(x)$. The electric quark current
$J^{\nu}(x)$ is defined as
\begin{eqnarray}\label{label1.9}
\hspace{-0.2in}J^{\nu}(x) = \sum_{i= q,\bar{q}} Q_i\int d \tau
\frac{d X^{\nu}_i(\tau)}{d\tau} \delta^{(4)}(x  - X_i(\tau)).
\end{eqnarray}
Hence, the inclusion of a dual Dirac string in terms of
 ${\cal E}^{\mu\nu}(x)$ defined by Eq.(\ref{label1.8}) satisfies completely
the electric Gauss law of Dirac$^{\prime}$s extension of Maxwell
$^{\prime}\,$s electrodynamics.

As has been shown in Refs.[1,2] the vacuum expectation values of
time--ordered products of densities expressed in terms of the
massless--monopole field, i.e., the magnetic monopole Green function
\begin{equation}\label{label1.10}
G\,(x_{1},\ldots, x_{n})=<{0}|{\rm T}(\bar{\chi} (x_{1}) \Gamma_{1}
\chi (x_{1}) \ldots \bar{\chi} (x_{n})\,\Gamma_{n}
 \chi\,(x_{n}))|{0}>_{\rm conn.}\,,
\end{equation}
where $\Gamma_i (i = 1,\ldots,n)$ are the Dirac matrices, are given by
[1,2]
\begin{eqnarray}\label{label1.11}
\hspace{-0.7in}&&G(x_{1},\ldots,x_{n})=<{0}|{\rm T}(\bar{\chi}(x_{1})
 \Gamma_1 \chi(x_{1}) \ldots \bar{\chi}(x_{n})
 \Gamma_n \chi(x_{n}))|{0}>_{\rm conn.} = \nonumber\\
\hspace{-0.7in}&&= ^{(M)}\!<{0}|{\rm T} \Big( \bar{\chi}_M(x_{1})
\Gamma_{1} \chi_M(x_{1}) \ldots \bar{\chi}_M(x_{n}) \Gamma_{n}
\chi_M(x_{n})\nonumber\\
\hspace{-0.7in}&&\times \; e^{\displaystyle i\int d^4x
 \{-g\bar{\chi}_M(x) \gamma^\nu \chi_M(x) C_{\nu}(x) -
 \kappa \bar{\chi}_M(x) \chi_M(x) \sigma(x) + {\cal L}_{\rm
int}[\sigma(x)]\}}\Big)|{0}>^{(M)}_{\rm conn.}.
\end{eqnarray}
Here $|{0}\!>^{(M)}$ is the wave-function of the nonperturbative vacuum of
the MNJL--model in the condensed phase and $|{0}\!>$ the wave-function of
the noncondensed perturbative vacuum.
${\cal L}_{\rm int}[\sigma(x)]$ describes self--interactions of the
$\sigma$--field:
\begin{equation}\label{label1.12}
{\cal L}_{\rm int}[\sigma(x)] = - \kappa\,M_{\sigma}\,
\sigma^3(x) - \frac{1}{2}\,\sigma^4(x).
\end{equation}
The self--interactions ${\cal L}_{\rm int}[\sigma(x)]$ provide
$\sigma$--field loop contributions and can be dropped out in the tree
$\sigma$--field approximation accepted in Refs. [1,2]. In the tree
$\sigma$--field
approximation the r.h.s. of Eq.(\ref{label1.11}) acquires the form
\begin{eqnarray}\label{label1.13}
\hspace{-0.3in}&&G(x_{1},\ldots,x_{n})=<{0}|{\rm T}(\bar{\chi}(x_{1})
 \Gamma_1 \chi(x_{1}) \ldots \bar{\chi}(x_{n}) \Gamma_n
 \chi(x_{n}))|{0}>_{\rm conn.} = \nonumber\\
\hspace{-0.3in}&&= ^{(M)}\!<{0}|{\rm T} \Big( \bar{\chi}_M(x_{1})
 \Gamma_{1} \chi_M(x_{1}) \ldots \bar{\chi}_M(x_{n}) \Gamma_{n}
\chi_M(x_{n})\nonumber\\
\hspace{-0.3in}&&\times \; e^{\displaystyle i\int d^4x
\Big\{-g\bar{\chi}_M(x) \gamma^\nu \chi_M(x) C_{\nu}(x) -
\kappa \bar{\chi}_M(x) \chi_M(x)
 \sigma(x)\}}\Big)|{0}>^{(M)}_{\rm conn.}.
\end{eqnarray}
The tree $\sigma$--field approximation can be justified keeping massive
magnetic monopoles very heavy, i.e. $M \gg M_C$. This corresponds to the
 London limit $M_{\sigma} = 2\,M \gg M_C$ in the dual Higgs model with dual
Dirac strings [11].

In Ref.[2]  Eq.(\ref{label1.13}) has been applied to the computation of the
magnetic monopole condensate $<\!\bar{\chi}(x)\chi(x);{\cal E}\!>$ in
dependence of the dual Dirac string shape represented by the electric
string tensor
${\cal E}^{\mu\nu}(x)$  Eq.(\ref{label1.8}). The magnetic monopole
condensate $<\!\bar{\chi}(x)\chi(x);{\cal E}\!>$ has been calculated in the
tree $\sigma$--field approximation neglecting the fluctuations of the
dual--vector field $C_{\mu}$ around the Abrikosov flux line which satisfies
the equation
\begin{eqnarray}\label{label1.14}
(\Box + M^2_C) C^{\nu}[{\cal E}(x)] = -
\partial_{\mu}{^*{\cal E}}^{\mu\nu}(x),
\end{eqnarray}
and takes the form
\begin{eqnarray}\label{label1.15}
C^{\nu}[{\cal E}(x)]= - \int d^4x^{\prime} \Delta (x - x^{\prime}, M_C)
\partial^{\prime}_{\mu} {^*{\cal E}}^{\mu\nu} (x^{\prime}\,),
\end{eqnarray}
where $\Delta(x - x^{\prime}, M_C)$ is the Green function
\begin{eqnarray}\label{label1.16}
\Delta (x - x^{\prime}, M_C) =
\int\frac{d^4k}{(2\pi)^4}\frac{e^{\displaystyle
-ik\cdot(x-x^{\prime}\,)}}{M^2_C - k^2 - i 0}.
\end{eqnarray}
In this paper we will calculate the magnetic monopole condensate
$<\!\bar{\chi}(x)\chi(x);{\cal E}\!>$ in the tree $\sigma$--field
approximation but taking into account quantum fluctuations of the
dual--vector field
 $C_{\mu}$ around the Abriksov flux line $C^{\nu}[{\cal E}(x)]$. An
important role of such fluctuations for the formation of the interquark
potential has
 been pointed out in Ref. [12] within a dual Higgs model with dual Dirac
strings.

This paper is organized as follows: In Sect.2 we calculate the magnetic
monopole condensate in the tree $\sigma$--field approximation and
explicitly integrate out quantum fluctuations of the dual--vector field
$C_{\mu}$ around the Abrikosov flux line. In Sect.3 we calculate the
contibution of the string shape fluctuations to the magnetic monopole
condensate. In the Conclusion we discuss the obtained results.

\section{Quantum dual--vector field fluctuations}
\setcounter{equation}{0}

\hspace{0.2in} In the tree $\sigma$--field approximation we determine the
magnetic monopole condensate $<\!\bar{\chi}(x)\chi(x);{\cal E}\!>$
following [1] as
\begin{eqnarray}\label{label2.1}
\hspace{-0.5in}&&<\bar{\chi}(x)\chi(x);{\cal E}> - <\bar{\chi}(0)\chi(0)> =
^{(M)}\!<{0}|{\rm T} \Big( \bar{\chi}_M(x)\chi_M(x)\nonumber\\
\hspace{-0.5in}&&\times \; e^{\displaystyle i\int d^4z
\Big\{-g\bar{\chi}_M(z) \gamma^\nu \chi_M(z) C_{\nu}(z) -
\kappa \bar{\chi}_M(z) \chi_M(z)
\sigma(z)\}}\Big)|{0}>^{(M)}_{\rm conn.},
\end{eqnarray}
where the r.h.s. of Eq.(\ref{label2.1}) should vanish at
$C_{\mu} = \sigma = 0$.

The time ordering operator and vacuum wave--function act on the massive
magnetic monopole fields $\chi_M$ and the fields of collective excitations
$\sigma$ and $C_{\mu}$.

The calculation of vacuum expectation values of time--ordered products of
the dual--vector fields is convenient to perform by means of the path
integral method
\begin{eqnarray}\label{label2.2}
\hspace{-0.3in}&&<\bar{\chi}(x)\chi(x);{\cal E}> -
<\bar{\chi}(0)\chi(0)> = \frac{1}{Z}\int {\cal D}C_{\mu}
 e^{\displaystyle i\int d^4z\,{\cal L}_{\rm eff}[C_{\mu}(z)]}\,
^{(M)}\!<{0}|{\rm T} \Big( \bar{\chi}_M(x)\chi_M(x)\nonumber\\
\hspace{-0.3in}&&\times \; e^{\displaystyle i\int d^4z
\Big\{-g\bar{\chi}_M(z) \gamma^\nu \chi_M(z) C_{\nu}(z) -
\kappa \bar{\chi}_M(z) \chi_M(z)
\sigma(z)\}}\Big)|{0}>^{(M)}_{\rm conn.},
\end{eqnarray}
where $Z$ is a normalization factor determined as
\begin{eqnarray}\label{label2.3}
\hspace{-0.5in}Z = \int {\cal D}C_{\mu} e^{\displaystyle i\int d^4z\,
{\cal L}_{\rm eff}[C_{\mu}(z)]}.
\end{eqnarray}
The effective Lagrangian ${\cal L}_{\rm eff}[C_{\mu}(z)]$ is defined by
the part of the effective Lagrangian Eq.(\ref{label1.5}) related to the
$C_{\mu}$--field:
\begin{eqnarray}\label{label2.4}
{\cal L}_{\rm eff}[C_{\mu}(z)] =\frac{1}{4} F_{\mu\nu}(z)
F^{\mu\nu}(z) + \frac{1}{2} M^2_C C_{\mu}(z) C^{\mu}(z).
\end{eqnarray}
In order to integrate out quantum fluctuations of the dual--vector field
$C_{\mu}$ around the shape of the Abrikosov flux line we split the
$C_{\mu}$--field into a classical field $C_{\mu}[{\cal E}(z)]$ induced by
the Dirac string and quantum fluctuations $c_{\mu}(z)$ around that
classical background.  [12]:
\begin{eqnarray}\label{label2.5}
C_{\mu}(z) = C_{\mu}[{\cal E}(z)] + c_{\mu}(z),
\end{eqnarray}
where $C_{\mu}[{\cal E}(z)]$ satisfies Eq.(\ref{label1.14}), and
$c_{\mu}(z)$ are the fluctuations of the dual--vector field  having a
vanishing vacuum expectation value $<c_{\mu}(z)> = 0$. Substituting the
decomposition Eq.(\ref{label2.5}) in the Lagrangian Eq.(\ref{label2.4})
we arrive at the Lagrangian of the quantum fields $c_{\mu}(x)$ fluctuating
around the Abrikosov flux line.
\begin{eqnarray}\label{label2.6}
{\cal L}_{\rm eff}[C_{\mu}(z)] ={\cal L}_{\rm string}(z)
+\frac{1}{2}\,c_{\mu}(z)\Big[(\Box + M^2_C)\,g^{\mu\nu} -
\partial^{\mu} \partial^{\nu}\Big]\,c_{\nu}(z),
\end{eqnarray}
where we have used Eq.(\ref{label1.14}). The Lagrangian of the dual
 Dirac string ${\cal L}_{\rm string}(z)$ is defined [5,11--13]
\begin{eqnarray}\label{label2.7}
\int d^4z{\cal L}_{\rm string}(z) = \frac{1}{4}\,M^2_C\int \int d^4z d^4y
{\cal E}_{\mu\alpha}(z)
\Delta^{\alpha}_{\nu}(z - y, M_C){\cal E}^{\mu\nu}(y),
\end{eqnarray}
where $\Delta^{\alpha}_{\nu}(z - y, M_C) = (g^{\alpha}_{\nu} +
2\partial^{\alpha}\partial_{\nu}/M^2_C) \Delta (z - y;M_C)$.

Since the Lagrangian Eq.(\ref{label2.5}) is Gaussian with respect to the
$c_{\mu}$--field, we are able to integrate out the $c_{\mu}$--field
exactly.

Integrating over the $c_{\mu}$--field we reduce
$<\bar{\chi}(x)\chi(x);{\cal E}>$ to the form:
\begin{eqnarray}\label{label2.8}
\hspace{-0.5in}&&<\bar{\chi}(x)\chi(x);{\cal E}> -
<\bar{\chi}(0)\chi(0)> =  ^{(M)}\!<{0}|{\rm T}
\Big( \bar{\chi}_M(x)\chi_M(x)\,\nonumber\\
\hspace{-0.5in}&&\times \; e^{\displaystyle - i\,
\frac{1}{2}\,g^2 \int\!\!\!\int  d^4z d^4y\,[\bar{\chi}_M(z)
\gamma^\mu \chi_M(z)]\,D_{\mu\nu}(z - z^{\prime},
M_C)\,[\bar{\chi}_M(z^{\prime}\,) \gamma^\nu \chi_M(z^{\prime}\,)]}
\nonumber\\
\hspace{-0.5in}&&\times \; e^{\displaystyle i\int d^4z
 \{-g\bar{\chi}_M(z) \gamma^\nu \chi_M(z) C_{\nu}[{\cal E}(z)] -
\kappa \bar{\chi}_M(z) \chi_M(z)
\sigma(z)\}}\Big)|{0}>^{(M)}_{\rm conn.},
\end{eqnarray}
where $D_{\mu\nu}(z - z^{\prime}, M_C)$ is the Green function of the free
$c_{\mu}$--field: $D_{\mu\nu}(z - z^{\prime}, M_C) = (g_{\mu\nu} +
\partial_{\mu}\partial_{\nu}/M^2_C)\,\Delta (z - z^{\prime}, M_C)$.
Since herein we consider dual Dirac strings as classical objects the
contribution of the Lagrangian of the dual Dirac strings ${\cal L}_{\rm
string}(z)$ cancels out.

The integration over $\sigma$--field degrees of freedom in the tree
approximation we perform following [2]. This yields
\begin{eqnarray}\label{label2.9}
\hspace{-0.5in}&&<\bar{\chi}(x)\chi(x);{\cal E}> -
<\bar{\chi}(0)\chi(0)> =  -  \frac{\kappa^2}{4M^3}\,
<\bar{\chi}(0)\chi(0)>\,^{(M)}\!<{0}|{\rm T}
\Big( \bar{\chi}_M(x)\chi_M(x) \nonumber\\
\hspace{-0.5in}&&\times \; e^{\displaystyle -
i \frac{1}{2} g^2 \int\!\!\!\int  d^4z d^4z^{\prime}
[\bar{\chi}_M(z) \gamma^\mu \chi_M(z)]\,
D_{\mu\nu}(z - z^{\prime}; M_C)\,[\bar{\chi}_M(z^{\prime}\,)
\gamma^\nu \chi_M(z^{\prime}\,)]}\nonumber\\
\hspace{-0.5in}&&\times \; e^{\displaystyle -i\,g \int d^4z \,
[\bar{\chi}_M(z) \gamma^\nu \chi_M(z)]\, C_{\nu}
[{\cal E}(z)]}\Big)|{0}>^{(M)}_{\rm conn.},
\end{eqnarray}
For the calculation of the vacuum expectation value in the r.h.s of
Eq.(\ref{label2.9}) we assume that the massive magnetic monopole fields
$\chi_M(x)$ are almost on--mass shell. It is valid due to a very large
mass of the  monopole fields. In this case the transfer momenta are small
compared with the mass of the dual--vector field $M_C$. By virtue these
assumptions we can reduce the four--monopole interaction in
Eq.(\ref{label2.9}) to a point--like interaction.
\begin{eqnarray}\label{label2.10}
\hspace{-0.5in}&&<\bar{\chi}(x)\chi(x);{\cal E}> -
<\bar{\chi}(0)\chi(0)> =  -  \frac{\kappa^2}{4M^3}\,
<\bar{\chi}(0)\chi(0)>\,^{(M)}\!<0|{\rm T}
\Big( \bar{\chi}_M(x)\chi_M(x) \nonumber\\
\hspace{-0.5in}&&\times \; e^{\displaystyle -
i\int d^4z\Bigg\{ \frac{g^2 }{2M^2_C}[\bar{\chi}_M(z)
\gamma^\mu \chi_M(z)]^2 + g \,[\bar{\chi}_M(z)
\gamma^\nu \chi_M(z)]\,
 C_{\nu}[{\cal E}(z)]\Bigg\}}\Big)|0>^{(M)}_{\rm conn.}.
\end{eqnarray}
Thus, since $M\gg M_C$ the vacuum averaging over the massive magnetic
monopole fields can be represented by the momentum integrals [2] related to
the magnetic monopole diagrams depicted in Fig. 1 and Fig. 2:
\begin{eqnarray}\label{label2.11}
\hspace{-0.5in}&&^{(M)}\!<0|{\rm T}
\Big( \bar{\chi}_M(x)\chi_M(x)\nonumber\\
\hspace{-0.5in}&&\times \; e^{\displaystyle -
i\int d^4z\Bigg\{ \frac{g^2 }{2M^2_C}[\bar{\chi}_M(z)
 \gamma^\mu \chi_M(z)]^2 + g \,[\bar{\chi}_M(z) \gamma^\nu
 \chi_M(z)]\,
 C_{\nu}[{\cal E}(z)]\Bigg\}}\Big)|0>^{(M)}_{\rm conn.}=\nonumber\\
\hspace{-0.5in}&&= - \frac{1}{16\pi^2}\int\frac{d^4k}{\pi^2i}{\rm
tr}\Bigg\{\frac{1}{M - \hat{k} + g\hat{C}[{\cal E}(x)]}-
\frac{1}{M - \hat{k}}\Bigg\} \nonumber\\
\hspace{-0.5in}&& - \frac{1}{16\pi^2}
\int\frac{d^4k}{\pi^2i}{\rm tr}\Bigg\{
\frac{1}{M - \hat{k} + g\hat{C}[{\cal E}(x)]}
\frac{1}{M - \hat{k} + g\hat{C}[{\cal E}(x)]}
\gamma_{\mu_1}\Bigg\}\nonumber\\
\hspace{-0.5in}&&\times \sum^{\infty}_{n=1}
\Bigg(\frac{g^2}{2M^2_C}\Bigg)^n
\Bigg(\frac{1}{16\pi^2}\Bigg)^{(n-1)}
\int\frac{d^4k_1}{\pi^2i}{\rm tr}\Bigg\{
\frac{1}{M - \hat{k}_1 + g\hat{C}[{\cal E}(x)]}
\gamma_{\mu_1}\frac{1}{M - \hat{k}_1 + g\hat{C}[{\cal
E}(x)]}\gamma_{\mu_2}\Bigg\}\nonumber\\
\hspace{-0.5in}&&\ldots \int\frac{d^4k_{n-1}}{\pi^2i}
{\rm tr}\Bigg\{\frac{1}{M - \hat{k}_{n-1} + g\hat{C}[{\cal
E}(x)]}\gamma_{\mu_{n-1}}\frac{1}{M - \hat{k}_{n-1} + g\hat{C}[{\cal
E}(x)]}\gamma_{\mu_n}\Bigg\}\nonumber\\
\hspace{-0.5in}&&\times \; \frac{1}{16\pi^2}
\int\frac{d^4k_n}{\pi^2i}{\rm tr}\Bigg\{\gamma_{\mu_n}
\frac{1}{M - \hat{k}_n + g\hat{C}[{\cal E}(x)]}\Bigg\} + \dots.
\end{eqnarray}
The first term in the r.h.s. of Eq.(\ref{label2.11}) has been calculated in
Ref.[2] at the neglect of quantum fluctuations of the dual--vector field
$C_{\mu}$, whereas the second term is fully due to these fluctuations. We
calculate the second term keeping leading divergent contributions as it is
accepted in the MNJL--model [1,2]. The ellipses denote the contribution of
the diagrams depicted  in Fig. 2b. This contribution is less important with
respect to the contribution of the diagrams in Fig. 2a and below we adduce
the result of the calculationof these diagrams without comments. The vacuum
expectation value Eq.(\ref{label2.11}) amounts to
\begin{figure}
\centerline{\epsfxsize=14cm \epsfbox{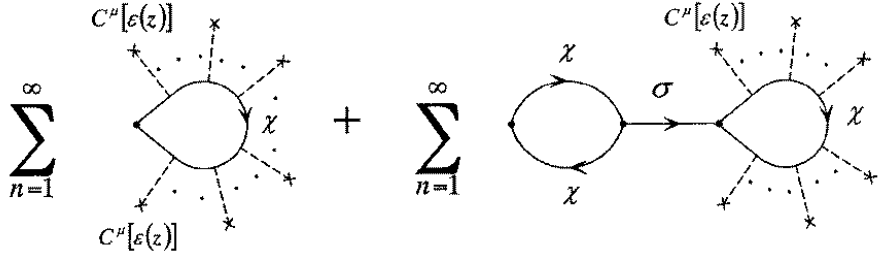}}
\caption{\label{Fig1}Magnetic monopole diagrams describing the magnetic
 monopole condensate around the dual Dirac string without fluctuations of
the dual--vector field.}
\end{figure}
\begin{figure}
\centerline{\epsfxsize=14cm \epsfbox{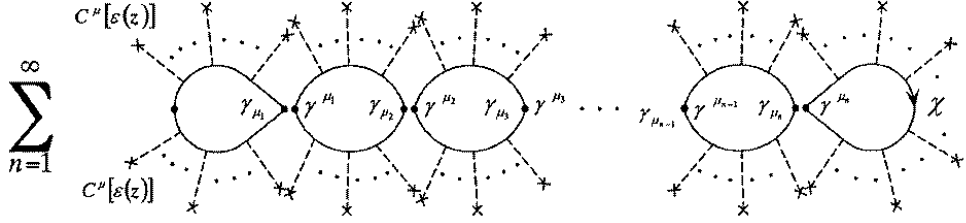}}
\centerline{(a)}
\vspace{.5cm}
\centerline{\epsfxsize=14cm \epsfbox{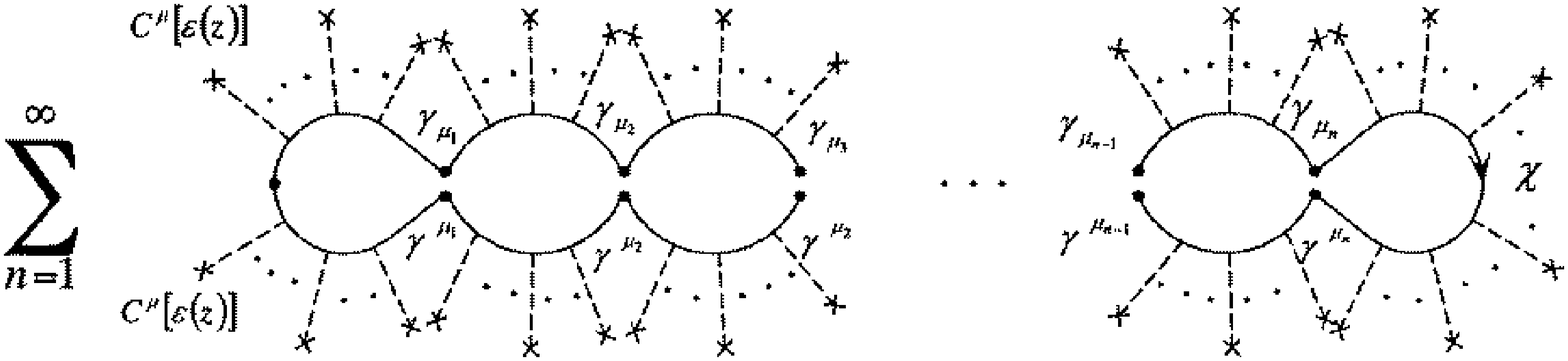}}
\centerline{(b)}
\caption{\label{Fig2}Magnetic monopole diagramms describing the
contributions to the magnetic monopole condensate caused by the
quantum fluctuations of the $C_{\mu}$--field.}
\end{figure}
\begin{eqnarray}\label{label2.12}
\hspace{-0.5in}&&^{(M)}\!<0|{\rm T}
\Big( \bar{\chi}_M(x)\chi_M(x)\nonumber\\
\hspace{-0.5in}&&\times\; e^{\displaystyle -
i\int d^4z\Bigg\{ \frac{g^2 }{2M^2_C}[\bar{\chi}_M(z)
\gamma^\mu \chi_M(z)]^2 + g \,[\bar{\chi}_M(z)
\gamma^\nu \chi_M(z)]\,
C_{\nu}[{\cal E}(z)]\Bigg\}}\Big)|0>^{(M)}_{\rm conn.}=
\nonumber\\
\hspace{-0.5in}&&= - \frac{M}{8\pi^2}\,
g^2\,C_{\mu}[{\cal E}(x)]C^{\mu}[{\cal E}(x)] +
\frac{M}{4\pi^2}\Bigg\{-\frac{g^2}{8\pi^2}
[J_1(M) + M^2J_2(M)]\Bigg\}\nonumber\\
\hspace{-0.5in}&&\times \; \frac{g^2}{2
M^2_C}\sum^{\infty}_{n=1}\Bigg\{\frac{1}{M^2_C}
\frac{g^2}{16\pi^2}[J_1(M) + M^2J_2(M)]\Bigg\}^{(n-1)}
C_{\mu}[{\cal E}(x)]C^{\mu}[{\cal E}(x)]\, =\nonumber\\
\hspace{-0.5in}&&= - \frac{M}{24\pi^2}\,
\frac{\displaystyle M^2_C +
\frac{g^2}{2G_1}}{\displaystyle M^2_C -
\frac{g^2}{6G_1}}\,g^2\,C_{\mu}[{\cal E}(x)]C^{\mu}[{\cal E}(x)].
\end{eqnarray}
Here we have used the definitions of $M^2_C$ given by Eq.(\ref{label1.2}).

Thus, integrating out explicitly quantum fluctuations of the dual--vector
field $C_{\mu}$ around the Abrikosov flux line and taking into account the
contribution of the scalar field $\sigma$ in the tree approximation we
obtain the magnetic monopole condensate in dependence on the shape of a
dual Dirac string in the form:
\begin{eqnarray}\label{label2.13}
\hspace{-0.5in}&&<\bar{\chi}(x)\chi(x);{\cal E}> =
<\bar{\chi}(0)\chi(0)>\nonumber\\
\hspace{-0.5in}&&\times \; \Bigg\{1 -
\frac{\kappa^2 G_1}{1024\pi^4}\,
 \frac{<\bar{\chi}(0)\chi(0)>}{M^3} +
\frac{\displaystyle M^2_C +
\frac{g^2}{2G_1}}{\displaystyle M^2_C -
\frac{g^2}{6G_1}}\,\frac{\kappa^2}{96\pi^2}\,
\frac{1}{M^2}\,g^2\,C_{\mu}[{\cal E}(x)]C^{\mu}[{\cal E}(x)]\Bigg\}.
\end{eqnarray}
The second term in the braces of Eq.(\ref{label2.13}) is the result of the
calculation of the diagrams in Fig. 2b.  It may be seen that quantum
fluctuations of a dual--vector field around the Abrikosov flux line give a
substantial contribution to the magnetic monopole condensate. In order to
retain the agreement with the results obtained within CQED [14] which
testify the suppression of the magnetic monopole condensate in the region
close to a dual Dirac string we have to impose the constraint $M^2_C >
g^2/6G_1$.
\vspace{1in}

\section{Dual Dirac string shape fluctuations}
\setcounter{equation}{0}

The string shape fluctuations we define following [13,15] by
$X^{\mu} \to X^{\mu} + \eta^{\mu}(X)$, where $\eta^{\mu}(X)$ describes
fluctuations around the fixed surface $S$ swept by the shape $\Gamma$ and
obeys the constraint $\eta^{\mu}(X)|_{\partial S} = 0$ [13,15] at the
boundary $\partial S$ of the surface $S$.

The magnetic monopole condensate defined by Eq.(\ref{label2.13}) and
averaged over string shape fluctuations reads
\begin{eqnarray}\label{label3.1}
\hspace{-0.4in}&&<\bar{\chi}(x)\chi(x);{\cal E}> -
<\bar{\chi}(0)\chi(0)>\Bigg[1 - \frac{\kappa^2 G_1}{1024\pi^4}\,
\frac{<\bar{\chi}(0)\chi(0)>}{M^3}\Bigg] = \nonumber\\
\hspace{-0.4in}&&<\bar{\chi}(0)\chi(0)>\frac{\displaystyle M^2_C +
\frac{g^2}{2G_1}}{\displaystyle M^2_C -
\frac{g^2}{6G_1}}\,\frac{\kappa^2}{96\pi^2}\,\frac{1}{M^2}\,
\frac{1}{Z_{\rm shape}}
\int {\cal D} \eta_{\mu} \,e^{\displaystyle i\,
\delta {\cal S}_{\rm N}[\eta]}\,
g^2\,C_{\nu}\{\eta (x)\}C^{\nu}\{\eta (x)\},
\end{eqnarray}
where $Z_{\rm shape}$ is a normalization factor determined as
\begin{eqnarray}\label{label3.2}
\hspace{-0.5in}Z_{\rm shape} = \int {\cal D}\eta_{\mu}
e^{\displaystyle i\,\delta {\cal S}_{\rm N}[\eta]}
\end{eqnarray}
and $\delta {\cal S}_{\rm N}[\eta]$ has been calculated in Ref.[13]:
\begin{eqnarray}\label{label3.3}
\delta\,{\cal S}_{\rm N}[\eta] = \int\!\!\!\int d^4x\,d^4y\,
\eta_{\alpha}(x)\,{\displaystyle O}^{\alpha\beta} (x - y)\,
\eta_{\beta}(y) + \int d^4x\,\eta_{\alpha}(x)\,
{\displaystyle O}^{\alpha} (x).
\end{eqnarray}
The operators ${\displaystyle O^{\alpha\beta}(x - y)}$ and
${\displaystyle O^{\alpha}(x)}$ are given by
\begin{eqnarray}\label{label3.4}
\hspace{-0.4in}{\displaystyle O}^{\alpha\beta}(x - y) &=&
 ~~\frac{1}{2}\,\delta^{(4)}(x - y)\,{\cal
E}_{\mu\nu}(x)\,\frac{\partial^2}{\partial x_{\alpha}\partial x_{\beta}}\,
\Sigma^{\nu\mu}(x) \nonumber\\
&&+\frac{1}{4}\,M^2_C\,{\cal E}_{\mu\nu}(x)\,
\frac{\partial^2}{\partial x_{\alpha}\partial
x_{\beta}}\,{\Delta^{\nu}}_{\lambda}(x-y; M_C)\,{\cal E}^{\lambda\mu}(y)
\nonumber\\
&&-{\displaystyle {\overleftarrow{\frac{\partial}{\partial
x_{\mu}}}}}\,\delta^{(4)}(x-y)\,{\cal
E}_{\mu\nu}(x)\,\Sigma^{\alpha\beta}(x)\,
{\displaystyle {\overrightarrow{\frac{\partial}{\partial y_{\nu}}}}}
\nonumber\\
&&+ {\displaystyle {\overleftarrow{\frac{\partial}{\partial x_{\mu}}}}}\,
{\cal E}_{\mu\nu}(x)\,\frac{\partial}{\partial x_{\beta}}
\Sigma^{\nu\alpha}(x)\,\delta^{(4)}(x - y) \nonumber\\
&&- \frac{1}{2}\,M^2_C\,
{\displaystyle {\overleftarrow{\frac{\partial}{\partial x_{\mu}}}}}\,
{\cal E}_{\mu\nu}(x)\,\frac{\partial}{\partial
x_{\beta}}{\Delta^{\alpha}}_{\lambda}(x - y; M_C)\,
{\cal E}^{\lambda\nu}(y) \nonumber\\
&& + \delta^{(4)}(x - y)\,\frac{\partial}{\partial
x_{\alpha}}\Sigma^{\beta\mu}(x)\,{\cal E}_{\mu\nu}(x)\,
{\displaystyle {\overrightarrow{\frac{\partial}{\partial y_{\nu}}}}}
 \nonumber\\
&& + \frac{1}{2}\,M^2_C\,
{\displaystyle {\overleftarrow{\frac{\partial}{\partial x_{\mu}}}}}
{\cal E}_{\mu\nu}(x)\,\frac{\partial}{\partial
x_{\beta}}{\Delta^{\nu}}_{\lambda}(x - y; M_C)\,
{\cal E}^{\lambda\alpha}(y) \nonumber\\
&&- \frac{1}{4}\,M^2_C\,
{\displaystyle {\overleftarrow{\frac{\partial}{\partial x_{\mu}}}}}\,
{\cal E}_{\mu\lambda}(x)\,\Delta^{\alpha\beta}(x - y; M_C)\,{\cal
E}^{\lambda\nu}(y) \,
{\displaystyle {\overrightarrow{\frac{\partial}{\partial y^{\nu}}}}}
\nonumber\\
&& - \frac{1}{4}\,M^2_C\,
{\displaystyle {\overleftarrow{\frac{\partial}{\partial
x^{\mu}}}}}\,g^{\alpha\beta}\,{\cal E}^{\mu\lambda}(x)\,
\Delta_{\lambda\rho}(x - y; M_C)\,{\cal E}^{\rho\nu}(y) \,
{\displaystyle {\overrightarrow{\frac{\partial}{\partial y^{\nu}}}}}
\nonumber\\
&& + \frac{1}{4}\,M^2_C\,
{\displaystyle {\overleftarrow{\frac{\partial}{\partial x_{\mu}}}}}
\,{\cal E}^{\mu\lambda}(x)\,\Delta_{\lambda\beta}(x - y; M_C)\,{\cal
E}^{\alpha\nu}(y)\,
{\displaystyle {\overrightarrow{\frac{\partial}{\partial y^{\nu}}}}}
\nonumber\\
&& + \frac{1}{4}\,M^2_C\,
{\displaystyle {\overleftarrow{\frac{\partial}{\partial x^{\mu}}}}}
\,{\cal E}^{\mu\beta}(x)\,{\Delta^{\alpha}}_{\lambda}(x - y; M_C)\,
{\cal E}^{\lambda\nu}(y)\,{\displaystyle
{\overrightarrow{\frac{\partial}{\partial y^{\nu}}}}},
\nonumber\\
\hspace{-0.4in}{\displaystyle O}^{\alpha}(x)
&=&~~{\cal E}_{\mu\nu}(x)\,\frac{\partial}{\partial
x_{\alpha}}\Sigma^{\nu\mu}(x) + \frac{1}{2}\,
{\cal E}_{\mu\nu}(x)\,\frac{\partial}{\partial x_{\mu}}
\Sigma^{\nu\alpha}(x) - \frac{1}{2}\,{\cal
E}_{\mu\nu}(x)\,\frac{\partial}{\partial x_{\mu}}\Sigma^{\alpha\nu}(x),
\end{eqnarray}
where $\Sigma^{\nu\mu}(x)$ is determined by
\begin{eqnarray}\label{label3.5}
\Sigma^{\nu\mu}(x) = \frac{1}{2}\,M^2_C\,\int d^4z\,
{\Delta^{\mu}}_{\lambda}(x - z; M_C)\,{\cal E}^{\lambda\nu}(z).
\end{eqnarray}
Using Eqs.(\ref{label1.8}) and (\ref{label1.15}) we determine
$g^2\,C_{\mu}\{\eta (x)\}C^{\mu}\{\eta (x)\}$ as follows:
\begin{eqnarray}\label{label3.6}
\hspace{-0.5in}&&g^2\,C_{\mu}\{\eta (x)\}C^{\mu}\{\eta (x)\}
= \nonumber\\
\hspace{-0.5in}&&=g^2 Q^2\int\!\!\!\int
d{^*\sigma}^{\lambda\mu}(X) d{^*\sigma}_{\rho\mu}(Y)
 \frac{\partial}{\partial x^{\lambda}}\Delta (x - X - \eta(X))
\frac{\partial}{\partial x_{\rho}}\Delta (x - Y - \eta(Y)).
\end{eqnarray}
The changes of the surface elements $\sigma^{\lambda\mu}(X)$ and
$d\sigma_{\rho\mu}(Y)$
caused by the shifts $X\to X + \eta(X)$ and $Y\to Y +\eta(Y)$ have not been
taken into account in the r.h.s. of Eqs.(\ref{label3.6}), since they vanish
for the straight string. Indeed, the integration over the $\eta$--field we
perform following [13,14] for fluctuations around the shape of the static
straight string with the length $L$ tracing out the rectangular surface $S$
with the time--side $T$. In this case the electric field strength
${\cal E}_{\mu\nu}(x)$ does not depend on time and reads
\begin{eqnarray}\label{label3.7}
\vec{\cal E}(\vec{x}\,) = \vec{e}_z\,Q\,\delta (x)\,\delta (y)\,
\Bigg[\theta \Bigg(z -\frac{1}{2}\,L\Bigg) - \theta \Bigg(z +
\frac{1}{2}\,L\Bigg)\Bigg],
\end{eqnarray}
where at $\vec{X}_q = (0,0,\frac{1}{2}\,L)$ and
$\vec{X}_{\bar{q}} = (0,0,-\frac{1}{2}\,L)$ quark and antiquark are placed,
respectively. Then the unit vector $\vec{e}_z$ is directed along the
$z$--axis and $\theta (z)$ is the Heaviside--step--function. The field
 strength Eq.(\ref{label3.7}) induces the dual--vector potential
\begin{eqnarray}\label{label3.8}
<\vec{C}(\vec{x}\,)> = - \,i\,Q\,\int \frac{d^3k}{4\,\pi^3}\,
\frac{\vec{k} \times \vec{e}_z}{k_z}\,\frac{1}{M^2_C +
\vec{k}^{\,2}}\,\sin\Bigg(\frac{k_z L}{2}\Bigg)\,
e^{i\,\vec{k}\cdot\vec{x}}.
\end{eqnarray}
Allowing only fluctuations in the plane perpendicular to the string
world--sheet, i.e. setting $\eta_t(t,z) = \eta_z(t,z) = 0$ [13,14],
we arrive at the fluctuation action
$\delta\,{\cal S}_{\rm N}[\eta_x,\eta_y]$
\begin{eqnarray}\label{label3.9}
\hspace{-0.5in}&&\delta\,{\cal S}_{\rm N}[\eta_x,\eta_y] = \nonumber\\
\hspace{-0.5in}&&=\int\limits^{T/2}_{-T/2} dt\int\limits^{T/2}_{-T/2}
dt^{\prime}\int\limits^{L/2}_{-L/2} dz \int\limits^{L/2}_{-L/2}
dz^{\prime}\Bigg[\frac{\partial \eta_x(t,z)}{\partial t}\,
O_1(t, z| t^{\prime}, z^{\prime}\,)\,\frac{\partial
\eta_x(t^{\prime},z^{\prime}\,)}{\partial t^{\prime}}\nonumber\\
\hspace{-0.5in}&&- \frac{\partial \eta_x(t,z)}{\partial z}\,O_2(t, z|
t^{\prime}, z^{\prime}\,)\,\frac{\partial
\eta_x(t^{\prime},z^{\prime}\,)}{\partial z^{\prime}}+\eta_x(t,z)\,O_3(t,
z| t^{\prime}, z^{\prime}\,)\,\eta_x(t^{\prime},z^{\prime}\,) + (x
\leftrightarrow y)\Bigg],
\end{eqnarray}
where the operators $O_i$ ($i=1,2,3$) are defined by
\begin{eqnarray}\label{label3.10}
\hspace{-0.5in}&&O_1(t, z| t^{\prime}, z^{\prime}\,) = Q^2 \int
\frac{d^2k_{\perp}}{64\pi^4}\int\limits^{\infty}_{-\infty}
\!\!\!\int\limits^{\infty}_{-\infty} dk_0 dk_z\,
{\displaystyle  e^{ \displaystyle - i k_0(t - t^{\prime} ) +
i k_z (z - z^{\prime} )}}\frac{\displaystyle M^2_C + k^2_z +
\frac{1}{2}\, \vec{k}^{\,2}_{\perp}}{\displaystyle M^2_C - k^2_0 +
k^2_z + \vec{k}^{\,2}_{\perp}},\nonumber\\
\hspace{-0.5in}&&O_2(t, z| t^{\prime}, z^{\prime}\,) = Q^2 \int
\frac{d^2k_{\perp}}{64\pi^4}\int\limits^{\infty}_{-\infty}
\!\!\!\int\limits^{\infty}_{-\infty} dk_0 dk_z\,
{\displaystyle  e^{ \displaystyle - i k_0(t - t^{\prime} ) +
i k_z (z - z^{\prime} )}} \frac{\displaystyle M^2_C - k^2_0 +
\frac{1}{2}\, \vec{k}^{\,2}_{\perp}}{\displaystyle M^2_C - k^2_0 +
 k^2_z + \vec{k}^{\,2}_{\perp}},\nonumber\\
\hspace{-0.5in}&&O_3(t, z| t^{\prime}, z^{\prime}\,)
=\delta(t -t^{\prime}\,)\,\delta(z -z^{\prime}\,) \,Q^2 \int
\frac{d^2k_{\perp}\vec{k}^{\,2}_{\perp}}{16\pi^3}
\int\limits^{\infty}_{-\infty}\frac{dk_z}{k_z}\,
\sin\Bigg(\frac{k_zL}{2}\Bigg)\,\cos(k_z z) \,\nonumber\\
&&\times
\frac{M^2_C + k^2_z}{M^2_C + \vec{k}^{\,2}_{\perp} + k^2_z} - Q^2
\int
\frac{d^2k_{\perp}\vec{k}^{\,2}_{\perp}}{64\pi^4}\int\limits^{\infty}_{-\infty}
\!\!\!\int\limits^{\infty}_{-\infty}  d k_0d k_z\,
{\displaystyle  e^{ \displaystyle - i k_0(t - t^{\prime}\,) +
i k_z (z - z^{\prime} )}} \nonumber\\
&&\hspace{2in}\times
 \frac{M^2_C - k^2_0+ k^2_z}{M^2_C - k^2_0 +
\vec{k}^{\,2}_{\perp} + k^2_z}.
\end{eqnarray}
The linear terms in the $\eta$--field expansion do not appear,
 since only the components $\Sigma_{tz}(x)$ and $\Sigma_{zt}(x)$
survive in Eq.(\ref{label3.4})  for the static string strained along
the $z$--axis.

The fluctuating fields $\eta_i(t,z)$, where $i=x,y$, should obey the
 boundary conditions $\eta_i(t,z)|_{\partial S} = 0$, which for the
rectangular surface read [13,14]
\begin{eqnarray}\label{label3.11}
\eta_i(t,z)|_{\partial S} = \eta_i(\pm T/2,z) =
\eta_i(t,\pm L/2) =\eta_i(\pm T/2,\pm L/2) =0.
\end{eqnarray}
The integration over the $\eta$--fields should be performed with the weight
\begin{eqnarray}\label{label3.12}
\frac{1}{Z_{\rm shape}}\int\!\!\!\int {\cal D}\eta_x{\cal D}\eta_y
\,{\displaystyle e^{\displaystyle i\delta\,
{\cal S}_{\rm N}[\eta_x,\eta_y]}},
\end{eqnarray}
where the measure of the integration reads
\begin{eqnarray}\label{label3.13}
{\cal D}\eta_x{\cal D}\eta_y = \prod_{-T/2\le t \le T/2}
\prod_{-L/2 \le z \le L/2} d \eta_x(t,z)\,d \eta_y(t,z) .
\end{eqnarray}
Before the integration over the $\eta$--fields we can make some
simplifications of the $\Delta$--functions. For this aim we suggest to
integrate out $\vec{k}_{\perp}$ keeping only the main divergent
contributions as it is accepted in our effective approach [1,2]. In the
region $-L/2 \le z \le L/2$ this reduces the operators $O_i$ ($i=1,2,3$)
to the expressions
\begin{eqnarray}\label{label3.14}
&&O_1(t, z| t^{\prime}, z^{\prime}\,) =
O_2(t, z| t^{\prime}, z^{\prime}\,) =
\frac{Q^2\Lambda^2_{\perp}}{32\pi}\,\delta(t -t^{\prime}\,)\,
\delta(z -z^{\prime}\,),\nonumber\\
&&O_3(t, z| t^{\prime}, z^{\prime}\,)=\frac{Q^2\Lambda^2_{\perp}}{16\pi}\,
\Bigg( -\frac{\partial^2}{\partial t^2} + \frac{\partial^2}{\partial
z^2}\Bigg)\,\delta(t -t^{\prime}\,)\,\delta(z -z^{\prime}\,),
\end{eqnarray}
where $\Lambda_{\perp}$ is the cut--off in the plane perpendicular to the
world--sheet of the string. The fluctuation action becomes
\begin{eqnarray}\label{label3.15}
\hspace{-0.5in}\delta\,{\cal S}_{\rm N}[\eta_x,\eta_y] = -
\frac{3Q^2\Lambda^2_{\perp}}{32\pi}\int\limits^{T/2}_{-T/2} dt
\int\limits^{L/2}_{-L/2} dz [\eta_x(t,z)\,(- \Delta)\,\eta_x(t,z) +
 (x \leftrightarrow y)],
\end{eqnarray}
where $\Delta$ is the Laplace operator in 2--dimensional space--time
\begin{eqnarray}\label{label3.16}
\Delta = - \frac{\partial^2}{\partial t^2} +
\frac{\partial^2}{\partial z^2}
\end{eqnarray}
The common factor $Q^2\Lambda^2_{\perp}/8\pi$ can be removed by the
renormalization of the $\eta$--fields, and the action of the fluctuations
becomes
\begin{eqnarray}\label{label3.17}
\hspace{-0.5in}\delta\,{\cal S}_{\rm N}[\eta_x,\eta_y] = -
\int\limits^{T/2}_{-T/2} dt \int\limits^{L/2}_{-L/2} dz
\Bigg[\eta_x(t,z)\,\Bigg(- \frac{\Delta}{M^2_C}\Bigg)\,
\eta_x(t,z) + (x \leftrightarrow y)\Bigg].
\end{eqnarray}
The factor $1/M^2_C$ is introduced by dimensional considerations. We have
 used the mass of the dual--vector field, since the Abrikosov flux line is
localized in the region of order of $O(1/M_C)$ in the $xy$--plane.
Of course, the final result does not depend on the parameter making the
operator $\Delta$ dimensionless.

For a static dual Dirac string and after the renormalization of the
$\eta$--fields the scalar product
$g^2\,C_{\mu}\{\eta (x)\}C^{\mu}\{\eta (x)\}$ amounts to
\begin{eqnarray}\label{label3.18}
\hspace{-0.4in}&&g^2\,C_{\mu}\{\eta (x)\}C^{\mu}\{\eta (x)\} = g^2
Q^2\int\!\!\!\int \frac{d^3k}{4\pi^3}\frac{d^3q}{4\pi^3}
\frac{k_x q_x + k_y q_y}{k_z q_z}\sin\Bigg(\frac{k_zL}{2}\Bigg)
\,\sin\Bigg(\frac{q_zL}{2}\Bigg)\,\frac{1}{M^2_C + \vec{k}^{\,2}}\,
\nonumber\\
\hspace{-0.4in}&&\times \; \frac{1}{M^2_C + \vec{q}^{\,2}}
\;e^{\displaystyle i\,(\vec{k} + \vec{q})\cdot \vec{x}}
\;e^{\displaystyle i\,
\sqrt{\frac{8\pi}{3}}\frac{1}{Q}
\frac{2}{M_C\Lambda_{\perp}}[(k_x + q_x) \eta_x(t,z) +
 (k_y + q_y) \eta_y(t,z)]}.
\end{eqnarray}

Thus, in the static dual Dirac string approximation Eq.(\ref{label3.1})
 reads
\begin{eqnarray}\label{label3.19}
\hspace{-0.4in}&&<\bar{\chi}(x)\chi(x);{\cal E}>-
<\bar{\chi}(0)\chi(0)>\Bigg[1 - \frac{\kappa^2 G_1}{1024\pi^4}\,
\frac{<\bar{\chi}(0)\chi(0)>}{M^3}\Bigg] = \nonumber\\
\hspace{-0.4in}&&=<\bar{\chi}(0)\chi(0)>
\frac{\displaystyle M^2_C +
\frac{g^2}{2G_1}}{\displaystyle M^2_C -
\frac{g^2}{6G_1}}\,\frac{\kappa^2}{96\pi^2}\,
\frac{1}{M^2}\,\nonumber\\
\hspace{-0.4in}&&\times \; g^2 Q^2\int\!\!\!\int
\frac{d^3k}{4\pi^3}\frac{d^3q}{4\pi^3}
\frac{k_x q_x + k_y q_y}{k_z q_z}\sin\Bigg(\frac{k_zL}{2}\Bigg)
\,\sin\Bigg(\frac{q_zL}{2}\Bigg)\,\frac{1}{M^2_C + \vec{k}^{\,2}}\,
\frac{1}{M^2_C + \vec{q}^{\,2}}\,
e^{\displaystyle i\,(\vec{k} + \vec{q})\cdot \vec{x}}
\nonumber\\
\hspace{-0.4in}&&\times \; \frac{1}{Z_{\rm shape}}
\int {\cal D}\eta_x{\cal D}\eta_y\,
\exp - i\int\limits^{T/2}_{-T/2} dt^{\prime}
\int\limits^{L/2}_{-L/2} dz^{\prime}
\Bigg[\eta_x(t^{\prime},z^{\prime}\,)\,
\Bigg(- \frac{\Delta}{M^2_C}\Bigg)\,\eta_x(t^{\prime},z^{\prime}\,)
 \nonumber\\
\hspace{-0.4in}&& - \sqrt{\frac{8\pi}{3}}\frac{1}{Q}
\frac{2}{M_C\Lambda_{\perp}}\,(k_x + q_x) \delta(t - t^{\prime}\,)
\, \delta(z - z^{\prime}\,)\,\eta_x(t^{\prime},z^{\prime}\,)
 + (x \leftrightarrow y)\Bigg].
\end{eqnarray}
Integrating over the $\eta$--fields we get
\begin{eqnarray}\label{label3.20}
\hspace{-0.6in}&&<\bar{\chi}(x)\chi(x);{\cal E}>-
<\bar{\chi}(0)\chi(0)>\Bigg[1 - \frac{\kappa^2 G_1}{1024\pi^4}\,
\frac{<\bar{\chi}(0)\chi(0)>}{M^3}\Bigg] = \nonumber\\
\hspace{-0.6in}&&=<\bar{\chi}(0)\chi(0)>
\frac{\displaystyle M^2_C +
 \frac{g^2}{2G_1}}{\displaystyle M^2_C -
\frac{g^2}{6G_1}}\,\frac{\kappa^2}{96\pi^2}\,\frac{1}{M^2}\,\nonumber\\
\hspace{-0.6in}&&\times \; g^2 Q^2\int\!\!\!\int
\frac{d^3k}{4\pi^3}\frac{d^3q}{4\pi^3}
\frac{\vec{k}_{\perp}\cdot \vec{q}_{\perp}}{k_z
q_z}\sin\Bigg(\frac{k_zL}{2}\Bigg)
\,\sin\Bigg(\frac{q_zL}{2}\Bigg)\,\frac{1}{M^2_C +
\vec{k}^{\,2}}\,\frac{1}{M^2_C + \vec{q}^{\,2}}\,e^{\displaystyle
i\,(\vec{k} + \vec{q})\cdot \vec{x}}\nonumber\\
\hspace{-0.6in}&&\times \; \exp\Bigg\{\displaystyle -
i\frac{8\pi}{3}\frac{(\vec{k}_{\perp} +
\vec{q}_{\perp})^2}{Q^2\Lambda^2_{\perp}}
\int\limits^{\infty}_{-\infty}\!\! dt^{\prime}
 \int\limits^{L/2}_{-L/2}\!\!
 dz^{\prime} \delta (t - t^{\prime})\delta (z - z^{\prime}\,)
\Delta^{-1}\delta (t - t^{\prime})\delta (z - z^{\prime}\,)\Bigg\},
\end{eqnarray}
where $\vec{k}_{\perp}\cdot \vec{q}_{\perp} = k_x q_x + k_y q_y$.

In the integrand the Green function
$\Delta^{-1}\delta (t - t^{\prime})\,\delta (z - z^{\prime}\,)$
should be calculated at certain boundary conditions. For the open
dual Dirac string the calculations should be performed using Dirichlet
 boundary conditions [13,14]. Since in this case
$\delta (z - z^{\prime}\,)$ is given by
\begin{eqnarray}\label{label3.21}
\delta (z - z^{\prime}\,) = \frac{2}{L}\sum^{\infty}_{n=-\infty}
\sin\Bigg(\frac{2\pi n}{L}  z\Bigg)\,
\sin\Bigg(\frac{2\pi n}{L} z^{\prime}\,\Bigg),
\end{eqnarray}
the Green function $\Delta^{-1}\delta (t - t^{\prime})\,
\delta (z - z^{\prime}\,)$ is defined
\begin{eqnarray}\label{label3.22}
\Delta^{-1}\delta (t - t^{\prime})\,\delta (z - z^{\prime}\,) =
\frac{2}{L}\sum^{\infty}_{n=-\infty}
\int\limits^{\infty}_{-\infty}\frac{d\omega}{2\pi}
\frac{\displaystyle e^{\displaystyle -i\omega (t -t^{\prime}\,)}}
{\displaystyle \omega^2 - \frac{4\pi^2 n^2}{L^2}}
\sin\Bigg(\frac{2\pi n}{L}z\Bigg)\,
\sin\Bigg(\frac{2\pi n}{L}z^{\prime}\,\Bigg).
\end{eqnarray}
Using Eq.(\ref{label3.22}) we reduce Eq.(\ref{label3.20}) to
the expression
\begin{eqnarray}\label{label3.23}
\hspace{-0.4in}&&<\bar{\chi}(x)\chi(x);{\cal E}>-
<\bar{\chi}(0)\chi(0)>\Bigg[1 - \frac{\kappa^2 G_1}{1024\pi^4}\,
\frac{<\bar{\chi}(0)\chi(0)>}{M^3}\Bigg] = \nonumber\\
\hspace{-0.4in}&&=<\bar{\chi}(0)\chi(0)>
\frac{\displaystyle M^2_C +
\frac{g^2}{2G_1}}{\displaystyle M^2_C -
\frac{g^2}{6G_1}}\,\frac{\kappa^2}{96\pi^2}\,
\frac{1}{M^2}\,\nonumber\\
\hspace{-0.5in}&&\times \; g^2 Q^2\int\!\!\!\int
\frac{d^3k}{4\pi^3}\frac{d^3q}{4\pi^3}
\frac{\vec{k}_{\perp}\cdot \vec{q}_{\perp}}{k_z
q_z}\sin\Bigg(\frac{k_zL}{2}\Bigg) \,
\sin\Bigg(\frac{q_zL}{2}\Bigg)\,
\frac{1}{M^2_C + \vec{k}^{\,2}}\,\frac{1}{M^2_C +
\vec{q}^{\,2}}\,e^{\displaystyle i\,(\vec{k} + \vec{q})\cdot
\vec{x}}\nonumber\\
\hspace{-0.5in}&&\times \; \exp\Bigg\{\displaystyle -
i\frac{8\pi}{3}\frac{(\vec{k}_{\perp} +
\vec{q}_{\perp})^2}{Q^2\Lambda^2_{\perp}}\frac{2}{L}
\sum^{\infty}_{n=-\infty}\int\limits^{\infty}_{-\infty}
\frac{d\omega}{2\pi}\frac{1}{\displaystyle \omega^2 -
 \frac{4\pi^2 n^2}{L^2}}
\sin^2\Bigg(\frac{2\pi n}{L}z\Bigg)\Bigg\}.
\end{eqnarray}
By applying the Wick rotation $\omega \to i\omega$ we obtain the magnetic
monopole condensate in the form
\begin{eqnarray}\label{label3.24}
\hspace{-0.4in}&&<\bar{\chi}(x)\chi(x);{\cal E}>-
<\bar{\chi}(0)\chi(0)>\Bigg[1 - \frac{\kappa^2 G_1}{1024\pi^4}\,
\frac{<\bar{\chi}(0)\chi(0)>}{M^3}\Bigg] = \nonumber\\
\hspace{-0.4in}&&=<\bar{\chi}(0)\chi(0)>
\frac{\displaystyle M^2_C + \frac{g^2}{2G_1}}
{\displaystyle M^2_C - \frac{g^2}{6G_1}}\,
\frac{\kappa^2}{96\pi^2}\,\frac{1}{M^2}\,\nonumber\\
\hspace{-0.5in}&&\times \; \int\!\!\!\int
\frac{d^3k}{2\pi^2}\frac{d^3q}{2\pi^2}
\frac{\vec{k}_{\perp}\cdot \vec{q}_{\perp}}{k_z
q_z}\sin\Bigg(\frac{k_zL}{2}\Bigg) \,
\sin\Bigg(\frac{q_zL}{2}\Bigg)\,
\frac{1}{M^2_C + \vec{k}^{\,2}}\,
\frac{1}{M^2_C + \vec{q}^{\,2}}\,
e^{\displaystyle i\,(\vec{k} + \vec{q})\cdot \vec{x}}\nonumber\\
\hspace{-0.5in}&&\times \; \exp\Bigg\{\displaystyle - \frac{1}{2}
\frac{(\vec{k}_{\perp} +
\vec{q}_{\perp})^2}{\Lambda^2_{\perp}}\,\varphi(z)\Bigg\},
\end{eqnarray}
where we have used the Dirac quantization condition $g\,Q = 2\pi$ and denoted
\begin{eqnarray}\label{label3.25}
\hspace{-0.2in}\varphi(z) =\frac{4}{3}
\frac{g^2}{\pi}\,\frac{2}{L}\sum^{\infty}_{n=-\infty}
\int\limits^{\infty}_{-\infty}\frac{d\omega}{2\pi}
\frac{1}{\displaystyle \omega^2 + \frac{4\pi^2 n^2}{L^2}}
\sin^2\Bigg(\frac{2\pi n}{L}z\Bigg).
\end{eqnarray}
The function $\varphi(z)$ is defined by a divergent series. Therefore, it
should be regularized. The regularization of this function we perform in
the Appendix. As it is shown the regularized $\varphi(z)$--function equals
to zero for any
$z$ ranging the values from the interval $ - L/2 \le z L/2$. Thus, below we
set $\varphi(z) = 0$ and get
\begin{eqnarray}\label{label3.26}
\hspace{-0.4in}&&<\bar{\chi}(x)\chi(x);{\cal E}>-
<\bar{\chi}(0)\chi(0)>\Bigg[1 - \frac{\kappa^2 G_1}{1024\pi^4}\,
\frac{<\bar{\chi}(0)\chi(0)>}{M^3}\Bigg] = \nonumber\\
\hspace{-0.4in}&&=<\bar{\chi}(0)\chi(0)>\frac{\displaystyle M^2_C +
\frac{g^2}{2G_1}}{\displaystyle M^2_C -
\frac{g^2}{6G_1}}\,\frac{\kappa^2}{96\pi^2}\,\frac{1}{M^2}\,\nonumber\\
\hspace{-0.5in}&&\times \; \int\!\!\!\int
\frac{d^3k}{2\pi^2}\frac{d^3q}{2\pi^2}\frac{\vec{k}_{\perp}\cdot
\vec{q}_{\perp}}{k_z q_z}\sin\Bigg(\frac{k_zL}{2}\Bigg)
\,\sin\Bigg(\frac{q_zL}{2}\Bigg)\,\frac{1}{M^2_C + \vec{k}^{\,2}}\,
\frac{1}{M^2_C + \vec{q}^{\,2}}\,
e^{\displaystyle i\,(\vec{k} + \vec{q})\cdot \vec{x}}.
\end{eqnarray}
For a sufficiently long string the main contributions to the integrals over
$k_z$ and $q_z$ come from the momenta $|k_z| \sim 2/L$ and
 $|q_z| \sim 2/L$.  These values are small compared with $M^2_C$ and can be
neglected in the denominators. This reduces the r.h.s. of
 Eq.(\ref{label3.24}) to the form
\begin{eqnarray}\label{label3.27}
\hspace{-0.5in}&&<\bar{\chi}(x)\chi(x);{\cal E}>-
<\bar{\chi}(0)\chi(0)>\Bigg[1 - \frac{\kappa^2 G_1}{1024\pi^4}\,
\frac{<\bar{\chi}(0)\chi(0)>}{M^3}\Bigg] = \nonumber\\
\hspace{-0.5in}&&=<\bar{\chi}(0)\chi(0)>\frac{\displaystyle M^2_C +
\frac{g^2}{2G_1}}{\displaystyle M^2_C -
\frac{g^2}{6G_1}}\,\frac{\kappa^2}{96\pi^2}\,\frac{1}{M^2}\,
\Bigg[\int\limits^{\infty}_{-\infty}
\frac{dk_z}{k_z}\sin\Bigg(\frac{k_z L}{2}\Bigg)
\cos(k_z z)\Bigg]^2\nonumber\\
\hspace{-0.5in}&&\times \;\int\!\!\!\int
\frac{d^2k_{\perp}}{2\pi^2}\frac{d^2q_{\perp}}{2\pi^2}\,
\frac{\displaystyle (\vec{k}_{\perp}\cdot \vec{q}_{\perp})}
{(M^2_C + \vec{k}^{\,2}_{\perp})
(M^2_C + \vec{q}^{\,2}_{\perp})}\,
e^{\displaystyle i\,
(\vec{k}_{\perp} + \vec{q}_{\perp})\cdot \vec{x}_{\perp}}.
\end{eqnarray}
where $\vec{x}_{\perp} = (x,y)$. Taking into account  that $z$ is in the
interval $- L/2 \le z \le L/2$ we simplify Eq.(\ref{label3.26}) as follows
\begin{eqnarray}\label{label3.28}
\hspace{-1in}&&<\bar{\chi}(x)\chi(x);{\cal E}>-
<\bar{\chi}(0)\chi(0)>\Bigg[1 -
\frac{\kappa^2 G_1}{1024\pi^4}\,
\frac{<\bar{\chi}(0)\chi(0)>}{M^3}\Bigg] = \nonumber\\
\hspace{-1in}&&=<\bar{\chi}(0)\chi(0)>
\frac{\displaystyle M^2_C +
\frac{g^2}{2G_1}}{\displaystyle M^2_C -
\frac{g^2}{6G_1}}\,\frac{\kappa^2}{96\pi^2}\,\frac{1}{M^2}\int\!\!\!\int
\frac{d^2k_{\perp}}{2\pi}
\frac{d^2q_{\perp}}{2\pi}\,
\frac{\displaystyle (\vec{k}_{\perp}\cdot \vec{q}_{\perp})\,
e^{\displaystyle i\,(\vec{k}_{\perp} + \vec{q}_{\perp})\cdot
\vec{x}_{\perp}}}{[M^2_C + \vec{k}^{\,2}_{\perp}]
[M^2_C + \vec{q}^{\,2}_{\perp}]}
\end{eqnarray}
We can represent the r.h.s. of Eq.(\ref{label3.27}) in the more convenient
form
\begin{eqnarray}\label{label3.29}
\hspace{-1in}&&<\bar{\chi}(x)\chi(x);{\cal E}>-
<\bar{\chi}(0)\chi(0)>\Bigg[1 - \frac{\kappa^2 G_1}{1024\pi^4}\,
\frac{<\bar{\chi}(0)\chi(0)>}{M^3}\Bigg] = \nonumber\\
\hspace{-1in}&&= - <\bar{\chi}(0)\chi(0)>
\frac{\displaystyle M^2_C +
\frac{g^2}{2G_1}}{\displaystyle M^2_C -
\frac{g^2}{6G_1}}\,\frac{\kappa^2}{96\pi^2}\,
\frac{1}{M^2}\Bigg[
{\bigtriangledown}_{\displaystyle
\vec{x}_{\perp}}\int\frac{d^2k_{\perp}}{2\pi}\,
\frac{\displaystyle e^{\displaystyle i\,\vec{x}_{\perp}\cdot
\vec{k}_{\perp}}}{M^2_C + \vec{k}^{\,2}_{\perp}}\Bigg]^2,
\end{eqnarray}
where ${\bigtriangledown}_{\displaystyle \vec{x}_{\perp}}$ is the gradiant
 with respect to $\vec{x}_{\perp}$.

Integrating over directions of the vector $\vec{k}_{\perp}$ and taking the
gradient we get
\begin{eqnarray}\label{label3.30}
\hspace{-0.7in}&&<\bar{\chi}(x)\chi(x);{\cal E}>-
<\bar{\chi}(0)\chi(0)>\Bigg[1 - \frac{\kappa^2 G_1}{1024\pi^4}\,
\frac{<\bar{\chi}(0)\chi(0)>}{M^3}\Bigg] = \nonumber\\
\hspace{-0.7in}&&= - <\bar{\chi}(0)\chi(0)>\frac{\displaystyle M^2_C +
\frac{g^2}{2G_1}}{\displaystyle M^2_C -
\frac{g^2}{6G_1}}\,\frac{\kappa^2}{96\pi^2}\,\frac{1}{M^2}
\Bigg[\int^{\infty}_0
\frac{dk k^2 J_1(kr)}{M^2_C + k^2}\Bigg]^2,
\end{eqnarray}
where $J_1(u k)$ is a Bessel function and $r = |\vec{x}_{\perp}|$.
The integral over $k$ can be calculated explicitly and reads
\begin{eqnarray}\label{label3.31}
\int^{\infty}_0\frac{dk k^2 J_1(kr)}{M^2_C + k^2} =
\frac{2M^2_C}{r}\int^{\infty}_0\frac{dk k J_0(kr)}{(M^2_C + k^2)^2}
 = M_C K_1(M_Cr),
\end{eqnarray}
where $ K_1(M_Cr)$ is a McDonald function.

Thus, the magnetic condensate averaged over quantum dual--vector field and
string shape fluctuations reads
\begin{eqnarray}\label{label3.32}
\hspace{-0.7in}&&<\bar{\chi}(x)\chi(x);{\cal E}>
=\nonumber\\
\hspace{-0.7in}&&=<\bar{\chi}(0)\chi(0)>
\Bigg[1 - \frac{\kappa^2 G_1}{1024\pi^4}\,
 \frac{<\bar{\chi}(0)\chi(0)>}{M^3} -
\frac{\displaystyle M^2_C +
\frac{g^2}{2G_1}}{\displaystyle M^2_C -
\frac{g^2}{6G_1}}\,\frac{\kappa^2}{96\pi^2}\,
\frac{M^2_C}{M^2}\,K^2_1(M_Cr)\Bigg].
\end{eqnarray}
It may be  seen that due to the constraint $M^2_C > g^2/6G_1$ the magnetic
monopole condensate at distances close to the string $r \to 0$ becomes
suppressed. For $r \to 0$ the McDonald function $K_1(M_Cr)$ behaves like
$K_1(M_Cr) \to 1/M_Cr$. However, we have to emphasize that in such a model
like the MNJL model [1,2] and a dual Higgs model [11] the region of
distances close to the string is restricted by the constraint $r \ge
1/\Lambda_{\perp}$, where $\Lambda_{\perp}$ is the cut--off in plane
perperdicular to the world--sheet of a dual Dirac string [2,5,11--13]. Due
to Nambu [5] $1/\Lambda_{\perp}$ should be understood as a thickness of a
string. Following [5,11] this cut--off $\Lambda_{\perp}$ should be
identified with the mass of the $\sigma$--meson, , i.e. $\Lambda_{\perp} =
M_{\sigma} = 2\,M$. As has been shown in Ref.[11] this choice makes
next--to--leading order corrections in large $M_{\sigma}$ expansion to the
string tension logarithmically small compared with the leading order
contribution.  Thus, the McDonald function $K_1(M_Cr)$ is restricted from
above as $K_1(M_Cr) \le 2M/M_C$. Since the value of the condensate can be
either negative or zero, we can impose the constraint
\begin{eqnarray}\label{label3.33}
1 - \frac{\displaystyle M^2_C +
\frac{g^2}{2G_1}}{\displaystyle M^2_C -
\frac{g^2}{6G_1}}\,\frac{\kappa^2}{24\pi^2} \ge 0,
\end{eqnarray}
where we have neglected the contribution of the term of order $O(1/M^3)$.
Using the relation $\kappa^2 = 2\,g^2/3$ we bring up Eq.(\ref{label3.33})
to the form
\begin{eqnarray}\label{label3.34}
M^2_C \ge \frac{g^2}{6 G_1}\,\frac{\displaystyle 1 +
\frac{g^2}{12\pi^2}}{\displaystyle 1 - \frac{g^2}{36\pi^2}}.
\end{eqnarray}
This relation agrees with the inequality $M^2_C > g^2/6G_1$ for
any $g^2/36\pi^2 < 1$.

At distances far from the string $r \to \infty$ the contribution of the
string is exponentially suppressed as $e^{\textstyle - 2\,M_C r}$ due to
the Meissner effect, and the magnetic monopole condensate tends to the
magnitude of the order parameter, i.e. $<\bar{\chi}(0)\chi(0)>$.  A similar
influence of an electric flux tube, being an analogy to a dual Dirac string
in CQED, on the magnitude of the magnetic monopole condensate has been
observed within CQED [15].

\section{Conclusion}

The investigation of the magnetic monopole condensate around a dual
Dirac string has shown that the integration over quantum fluctuations of
the dual--vector field $C_{\mu}$ around the shape of the Abrikosov flux
line leads to a substantial non--positive defined contribution. The former
might change the result obtained in Ref.[2] concerning the suppression of
the magnetic monopole condensate at distances close to the dual Dirac
string. In order to retain this suppression obtained in CQED [14] we have
to impose the contraint $M^2_C > g^2/6G_1$. Since the coupling constant
$G_1$ is arbitrary, the  mass of the dual--vector field $M^2_C$ is left
arbitrary to the same extent. Due to the constraint $M^2_C > g^2/6G_1$ the
contribution of quantum dual--vector field fluctuations to the magnetic
monopole condensate decreases at distances far from the string, where the
influence of the string is exponentially suppressed due to the Meissner
effect. At infinitely large distances the magnitude of the magnetic
monopole condensate tends to the magnitude of the order parameter, i.e.,
$<\bar{\chi}(0)\chi(0)>$. The integration over string shape fluctuations
can be performed analytically only for the fluctuations around the  shape
of the static straight string of length $L$. The contribution of the string
shape fluctuations smoothes the suppression of the magnetic monopole
condensate at distances close to the string and retains the exponential
decrease at distances far from the string.

\section*{Appendix. Regularization of the $\varphi(z)$--function}

The function $\varphi(z)$ represented Eq.(\ref{label3.25}) is defined by a
divergent expression. Therefore, it is requested to regularize it. For the
regularization of  $\varphi(z)$ we introduce an arbitrary infra--red
parameter $\mu$ as follows
$$
 \varphi(z) \to \varphi(z)_{\rm R} = \frac{4}{3}
\,\frac{g^2}{\pi}\,\frac{2}{L}\sum^{\infty}_{n=-\infty}
\,\int\limits^{\infty}_{-\infty}
\,\frac{d\omega}{2\pi}\frac{1}{\displaystyle \omega^2 +
 \frac{4\pi^2 n^2}{L^2} + \mu^2}
\,\sin^2\Bigg(\frac{2\pi n}{L}z\Bigg). \eqno (A.1)
$$
The next step of the regularization is to apply the following integral
representation:
$$
 \varphi(z)_{\rm R} = \lim_{\mu \to 0}\,\frac{4}{3}
\,\frac{g^2}{\pi}\,\frac{2}{L}\sum^{\infty}_{n=0}
\,\int\limits^{\infty}_{-\infty}
\,\frac{dt}{2\pi}\,e^{\displaystyle -i 2\pi n
t/L}\int\limits^{\infty}_{-\infty}\int\limits^{\infty}_{-\infty}
\,\frac{d\omega dp_z}{2\pi}
\,\frac{e^{\displaystyle i p_z t}}{\displaystyle
 \omega^2 + p^2_z + \mu^2}[1 - \cos(2 p_z z)]. \eqno(A.2)
$$
It is easy to show that integrating over $t$ and $p_z$ we return to Eq.(A.1).

Then, it is convenient to decompose the integrals into two parts
$$
 \varphi(z)_{\rm R} = \varphi^{(1)}(z)_{\rm R} - \varphi^{(2)}(z)_{\rm
R},\eqno(A.3)
$$
where we have denoted
$$
\varphi^{(1)}(z)_{\rm R}= \frac{4}{3} \frac{g^2}{\pi}
\,\frac{2}{L}\sum^{\infty}_{n=0}
\,\int\limits^{\infty}_{-\infty}\frac{dt}{2\pi}
\,e^{\displaystyle -i 2\pi n t/L}\int\limits^{\infty}_{-\infty}
\,\int\limits^{\infty}_{-\infty}\frac{d\omega dp_z}{2\pi}
\,\frac{e^{\displaystyle i p_z t}}
{\displaystyle \omega^2 + p^2_z + \mu^2}, \eqno(A.4)
$$
$$
 \varphi^{(2)}(z)_{\rm R} = \frac{4}{3} \frac{g^2}{\pi}
\,\frac{2}{L}\sum^{\infty}_{n=0}\int\limits^{\infty}_{-\infty}
\,\frac{dt}{2\pi}\,e^{\displaystyle -i 2\pi n t/L}
\,\int\limits^{\infty}_{-\infty}
\,\int\limits^{\infty}_{-\infty}\frac{d\omega dp_z}{2\pi}
\,\frac{e^{\displaystyle i p_z t}}{\displaystyle \omega^2 +
 p^2_z + \mu^2}\cos(2 p_z z) =
$$
$$
= \frac{2}{3} \frac{g^2}{\pi}
\,\frac{2}{L}\sum^{\infty}_{n=0}\int\limits^{\infty}_{-\infty}
\,\frac{dt}{2\pi}\,e^{\displaystyle -i 2\pi n t/L}
\,\int\limits^{\infty}_{-\infty} \int\limits^{\infty}_{-\infty}
\,\frac{d\omega dp_z}{2\pi}\frac{e^{\displaystyle ip_z(t + 2z)} +
\,e^{\displaystyle ip_z(t - 2z)}}{\displaystyle \omega^2 + p^2_z +
\mu^2}.\eqno(A.5)
$$
Now let us perform a summation over index $n$ which gives
$$
 \varphi^{(1)}(z)_{\rm R}=  \frac{4}{3} \frac{g^2}{\pi}
\,\frac{2}{L}\int\limits^{\infty}_{-\infty}\frac{dt}{4\pi i}
\,\frac{e^{\displaystyle i \pi t/L}}{\displaystyle \sin(\pi
t/L)}\int\limits^{\infty}_{-\infty}\int\limits^{\infty}_{-\infty}
 \frac{d\omega dp_z}{2\pi}
\,\frac{e^{\displaystyle i p_z t}}{\displaystyle \omega^2 + p^2_z + \mu^2},
\eqno(A.6)
$$
$$
 \varphi^{(2)}(z)_{\rm R} =  \frac{2}{3} \frac{g^2}{\pi}
\,\frac{2}{L}\int\limits^{\infty}_{-\infty}\frac{dt}{4\pi i}
\,\frac{e^{\displaystyle i \pi t/L}}{\displaystyle \sin(\pi
t/L)}\int\limits^{\infty}_{-\infty}
\,\int\limits^{\infty}_{-\infty}\frac{d\omega dp_z}{2\pi}
 \frac{e^{\displaystyle ip_z(t + 2z)} + e^{\displaystyle ip_z(t -
2z)}}{\displaystyle \omega^2 + p^2_z + \mu^2}.\eqno(A.7)
$$
It is convenient to proceed to polar coordinates in the plane
$(\omega, p_z)$ and perform the integration over the azimuthal angle:
$$
 \varphi^{(1)}(z)_{\rm R}= \frac{4}{3} \frac{g^2}{\pi}
\,\frac{2}{L}\int\limits^{\infty}_{-\infty}\frac{dt}{4\pi i}
\,\frac{e^{\displaystyle i \pi t/L}}{\displaystyle \sin(\pi
t/L)}\int\limits^{\infty}_{0}
\frac{dp\,p\,J_0(pt)}{\displaystyle p^2  + \mu^2}, \eqno(A.8)
$$
$$
 \varphi^{(2)}(z)_{\rm R} = \frac{2}{3} \frac{g^2}{\pi}\,\frac{2}{L}
\,\int\limits^{\infty}_{-\infty}\frac{dt}{4\pi i}
\,\frac{e^{\displaystyle i \pi t/L}}{\displaystyle \sin(\pi t/L)}
\,\int\limits^{\infty}_{0}
 \frac{dp\,p}{\displaystyle p^2 + \mu^2}[J_0(p(t + 2z)) +
J_0(p(t-2z))],\eqno(A.9)
$$
where $J_0(x)$ is a Bessel function. Since the Bessel functions in the
integrands of Eqs.(A.8) and (A.9) are even under the transformation
$t\to - t$, the integrals become
$$
 \varphi^{(1)}(z)_{\rm R} = \frac{4}{3} \frac{g^2}{\pi}
\,\frac{2}{L}\int\limits^{\infty}_{-\infty}\frac{dt}{4\pi}
\,\int\limits^{\infty}_{0}
 \frac{dp\,p\,J_0(pt)}{\displaystyle p^2 + \mu^2}, \eqno(A.10)
$$
$$
 \varphi^{(2)}(z)_{\rm R} = \frac{2}{3} \frac{g^2}{\pi}\,
\frac{2}{L}\int\limits^{\infty}_{-\infty}\frac{dt}{4\pi}
\,\int\limits^{\infty}_{0}\frac{dp\,p}{\displaystyle p^2 + \mu^2}
[J_0(p(t + 2z)) + J_0(p(t-2z))].\eqno(A.11)
$$
The dependence of $z$ can be removed by the shifts $t + 2z \to t$ and
 $t - 2z \to t$:
$$
 \varphi^{(1)}(z)_{\rm R} = \varphi^{(2)}(z)_{\rm R} = \frac{4}{3}
 \frac{g^2}{\pi}\,\frac{2}{L}
 \int\limits^{\infty}_{-\infty}\frac{dt}{4\pi}\,\int\limits^{\infty}_{0}
 \frac{dp\,p\,J_0(pt)}{\displaystyle p^2 + \mu^2}, \eqno(A.12)
$$
As the integral over $t$ equals to
$$
 \int\limits^{\infty}_{-\infty} dt\,p\,J_0(pt) = 2,\eqno(A.13)
$$
the functions $\varphi^{(1)}(z)_{\rm R}$ and
$\varphi^{(2)}(z)_{\rm R}$ are defined by the integral over $p$:
$$
 \varphi^{(1)}(z)_{\rm R}= \varphi^{(2)}(z)_{\rm R} = \frac{2}{3}
  \frac{g^2}{\pi^2}\,\frac{2}{L}\int\limits^{\infty}_{0}
 \frac{dp}{\displaystyle p^2 + \mu^2}= \frac{g^2}{3 \pi}
\,\frac{2}{L \mu}.\eqno(A.13)
$$
Substituting Eq.(A.13) in Eq.(A.3) we get
$$
 \varphi(z)_{\rm R} = 0.\eqno(A.14)
$$
Thus, the regularized version of the $\varphi(z)$--function vanishes
for all $z \in [-L/2,L/2]$.

\end{document}